\documentstyle[12pt]{article}
\begin{document}

\vskip 1cm

\underline{\bf Title: } {\bf Response to "Reanalysis of the spectrum of the z=10 galaxy ISAAC/VLT
observations of a lensed galaxy at z=10.0" by Weatherley et
al. (astro-ph/0407150)} 

\vskip 0.5cm

\underline{\bf Author:} R. Pello (1), J. Richard (1), J.-F. Le
Borgne (1), D. Schaerer (2, 1)

\smallskip
\smallskip

(1) Laboratoire d'Astrophysique (UMR 5572), Observatoire
    Midi-Pyr\'en\'ees, 14 Avenue E. Belin, F-31400 Toulouse, France

(2) Observatoire de Gen\`eve,
51, Ch. des Maillettes, CH-1290 Sauverny, Switzerland

\vskip 0.5cm

\underline{\bf Abstract:}

In a recent posting of a submitted paper Weatherley et
al. (astro-ph/0407150) re-analyse
the spectroscopic data from our observations of Abell 1835 IR 1916, a galaxy
we claimed to be at redshift z$=$10.0 (Pello et al., 2004, A\&A, 416, L35).
They conclude that the emission line is not detected in their analysis.
Although their reduction procedure is accurate and robust, there are several
differences with ours. The most important one concerns the stacking of the
spectral frames. In the case of compact target sources [here: seeing-limited object,
$\sim$3 pixels FWHM in the spectral direction, on a 1" ($\sim$7 pixels) wide slit]
spectral shifts occur between individual frames. Our reduction corrects for
this effect using a nearby strong emission line from a compact reference object,
as described in Pello et al. However, Weatherley et al. neglect this
effect. This difference turns out to be crucial. E.g. adopting their
combination scheme as closely as possible but correcting for the spectral
drifts we recover the line detection. We conclude that more work is needed
to fully understand the origin of our differences. However, their non-detection
claim needs further investigation and cannot be uphold based on the arguments
presented in their original version of the paper. More information is beeing
posted on 

\begin{center}
{\tt
  http://webast.ast.obs-mip.fr/people/roser/z10$\_$spectroscopy$\_$discussion\/}
\end{center}

With this material we also invite other researchers to re-analyse independently
these ISAAC spectroscopic observations.

\end{document}